\documentclass[conference, letterpaper, onecolumn]{IEEEtran}
\IEEEoverridecommandlockouts

\usepackage[noadjust]{cite}
\usepackage{amsmath,amssymb,amsfonts,amsthm}
\usepackage{algorithmic}
\usepackage{graphicx}
\usepackage{textcomp}
\usepackage{xcolor}
\usepackage{dsfont}
\usepackage{hyperref}

\usepackage[capitalise]{cleveref}
\usepackage{mathtools}
\usepackage{flushend}
\usepackage{microtype}

\def\BibTeX{{\rm B\kern-.05em{\sc i\kern-.025em b}\kern-.08em
    T\kern-.1667em\lower.7ex\hbox{E}\kern-.125emX}}

\usepackage{preamble}

\begin{document}

\title{Necessity of Block Designs for Optimal Locally Private Distribution Estimation\\
\thanks{This work was supported by an Australian Government Research Training Program (RTP) Scholarship.}
\thanks{This is a preprint of a paper accepted for presentation at ITW 2025.}
}

\author{\IEEEauthorblockN{Abigail Gentle}
\IEEEauthorblockA{\textit{School of Computer Science} \\
\textit{University of Sydney}\\
{abigail.gentle@sydney.edu.au}}
}

\maketitle

\begin{abstract}
	Local differential privacy represents the gold standard for preserving the privacy of data before it leaves the device, and distribution estimation under this model has been well studied. Recently, protocols built upon balanced incomplete block designs were shown to achieve optimal error for this problem. However, it remained unknown whether other constructions could also be optimal. We resolve this question by proving that any protocol achieving optimal error must correspond to some balanced incomplete block design. This result, combined with prior work, completely characterises the set of optimal protocols for this problem. As a consequence, the protocols that achieve optimal error and optimal communication are only those based on symmetrical balanced incomplete block designs.
\end{abstract}


\section{Introduction}
\label{introduction}
Over the past decades, as data collection has grown exponentially, so too have privacy concerns. In response to this need differential privacy (DP)~\cite{dwork2006, dwork2013} has emerged as the gold standard for privacy-preserving data analysis, providing rigorous worst-case guarantees. As most of this data is collected in a distributed fashion, a natural point to apply privacy has become on the device, motivating massive advances in the field of local differential privacy (LDP) over the past years. 

\begin{definition}[Local differential privacy~\cite{kasiviswanathan2011a}]
	\label{defdp}
	For $\priv>0$, a randomised algorithm $\Pmech$ provides $\priv$-\textit{local differential privacy} if for all $i,j\in \dmain$
	\begin{equation}
		\max_{y\in\range}\frac{\Pr[Q(i)=y]}{\Pr[Q(j)=y]}\leq e^\priv.
	\end{equation}
	In other words, the probability of seeing any specific output $y$ does not vary much between any two inputs $i$ and $j$. 
\end{definition}

We consider private distribution estimation over discrete domains. Let $\mu$ be a discrete distribution over the domain $\dmain\defeq[\ab]\defeq\{ 1,\dots,\ab \}$, and let $X\defeq\{ x_{1},\dots,x_{n}\}$ be $n$ i.i.d.samples from $\mu$, each held by one of $n$ (possibly distributed) users. We are interested in \textit{locally private protocols for distribution estimation}, defined as follows.

\begin{definition}[LDP distribution estimation]
	An $\priv$-LDP protocol for distribution estimation comprises a local randomiser $\Pmech\colon\dmain\to\range$ that satisfies~\cref{defdp} and a ``debiasing'' function $\mathcal{D}\colon\range^\ns\to\mathbb{R}^\ab$ that converts private outputs into an estimate of the true distribution.
\end{definition}

The output space $\range\defeq[\oab]\defeq\{1,\dots,\oab\}$ need not be the same size as that of $\dmain$, and is frequently much larger~\cite{erlingsson2014,wang2016,ye2018a, park2024}. The \emph{communication complexity} of a protocol is the number of bits a user must send to the server after applying the local randomiser. It is apparent that this is at most $\lceil\log_2(\oab)\rceil$ bits.

As $\Pmech$ amounts to a procedure mapping inputs onto probability distributions, we can represent it by its underlying transition probability matrix (TPM) where $\Pmech_{ij}=\Pr[Y=i\mid X=j]$. We abuse notation slightly, and allow $\Pmech$ to refer to the local randomiser and the TPM interchangeably. The problem now becomes learning the true probabilities of $\dist_i$ given $\ns$ samples over the induced distribution
\begin{align}
\label{induceddist}
	\freqs &= \Pmech\dist.
\end{align}

We measure the \emph{utility}, or \emph{accuracy}, of a protocol by the expected error of the debiased estimate. In this work we will consider the expected $\ell_2^2$ risk, 
\begin{equation}
	\bEE{\ell_2^2(\hat{\dist},\dist)} = \bEE{\norm{\hat{\dist}-\dist}_2^2}
\end{equation}
where the expectation is taken over the input distribution and the local randomiser.

Protocols that share a common source of randomness with the server are called ``public-coin'', and can reduce communication overhead and increase accuracy~\cite{chen2023,acharya2019b}. We here consider only ``private-coin'' protocols, where there is no coordination between the server and the users. If all users implement the same local randomiser the protocol is called \emph{symmetric}. Symmetric protocols are the only type considered in this work. 

Recent work has demonstrated that a certain family of protocols based on balanced incomplete block designs (as per~\cref{defbibd}) achieve minimax optimality (as per~\cref{minimaxrisk})~\cite{park2024}; however, it remained unknown whether other constructions could also be optimal. In~\cref{optdpimpliesbibd} we resolve this by proving the converse: \emph{any locally private protocol for distribution estimation that achieves minimax optimality must have a transition probability matrix derived from a balanced incomplete block design.} This implies that the family of block-design randomised response protocols in~\cite{park2024} exactly characterises \emph{all} minimax optimal protocols.

Balanced incomplete block designs, otherwise known as block designs, or $\bvr$-configurations are well-studied combinatorial objects, whose connection to statistics and experiment design are well known~\cite{raghavarao1988,ryser1963}. Historically, these objects were first considered in the context of randomised response procedures~\cite{raghavarao1979,padgett2005}. They have since been used implicitly to construct optimal LDP protocols for distribution estimation in the works of~\cite{acharya2019a,feldman2022a}. This connection was made explicit in~\cite{park2024}.

To aid our discussion of these objects, we first introduce their basic properties.

\begin{definition}[{\cite[Chapter 8]{ryser1963}}]
\label{defbibd}
	A \emph{balanced incomplete block design} (BIBD), or a $\bvr$\emph{-configuration} is a generalisation of finite projective planes. Let $\dmain$ be a set of $\ab$ elements (a $\ab$-set) and let $\cB_1,\cB_2,\dots,\cB_\oab$ be $\oab$ distinct subsets (blocks) of $\dmain$. These subsets form a balanced incomplete block design provided that: 
	\begin{enumerate}
		\item Each block $\cB_i$ has $\lvert \cB_i\rvert=\ssz$ (call it a $\ssz$-subset of $\dmain$),
		\item each 2-subset of $\dmain$ is a subset of exactly $\sint$ of the sets $\cB_1,\cB_2,\dots \cB_\oab$, and
		\item the integers $\ab$, $\ssz$, and $\sint$ satisfy $0<\sint$ and $\ssz < v-1$.
	\end{enumerate}
	These conditions imply that each element of $\dmain$ is contained in exactly $\reps$ blocks.

	Our use of the letters $v$, $b$, and $r$ reflect their origin in experiment design, where we have \emph{varieties} assigned to \emph{blocks}, and each variety is \emph{replicated} $\reps$ times\footnote{Typically it is required that $\sint>0$ to exclude trivial designs, however we do not make this distinction here.}. 
	These constraints ensure that
	\begin{align}
		\reps(\ssz - 1) &=\sint(\ab-1)\\
		\oab\ssz &= \ab\reps
	\end{align}
	Now let $A$ be the incidence matrix of this configuration. That is, $A$ is a zero-one matrix with $A_{ij}=1$ if $j\in\cB_i$. The properties above imply that
	\begin{align}
		AJ &= \ssz J' \\
		A'A &= (\reps - \sint)I + \sint J,
	\end{align}
	$A'$ denotes the transpose of $A$, $J$ is the all-ones matrix of order $\ab$, $J'$ is the all-ones matrix of size $\oab\times\ab$, and $I$ is the identity matrix of order $\ab$.

	Conversely, if $A$ satisfies these properties, then it is the incidence matrix of a BIBD.
	
	If $b=v$ and $r=k$, we call it a \emph{symmetrical balanced incomplete block design} (SBIBD), or a $\vkl$\textit{-configuration}.
\end{definition}

Optimal protocols for locally private distribution estimation are those that achieve the smallest error over all possible input distributions. We formalise this with the standard definition of minimax optimality.
\begin{definition}[Minimax optimality]
	\label{defminimax}
	A protocol $(\Pmech,\debias)$ is \emph{minimax optimal} for privacy level $\priv$, and for risk function $\risk$, if it achieves the minimax risk:
	\begin{equation}
	\sup_{\dist\in\Delta_\ab} \risk(\Pmech,\debias;\dist) = \inf_{\Pmech'\in\mathcal{C}}\inf_{\debias'}\sup_{\dist\in\Delta_\ab} \risk(\Pmech',\debias';\dist)
	\end{equation}
	where $\mathcal{C}$ is the set of all $\priv$-LDP local randomisers. In this work $\risk$ is the expected $\ell_2^2$ error.
\end{definition}

Our primary focus are protocols that are minimax, or ``exactly,'' optimal with respect to constant factors. This is in contrast with those protocols that are asymptotically optimal, i.e. achieving $O(\cdot)$ of the minimax optimal error. This is well motivated by the statistical nature of the problem discussed, where collecting twice as many samples (let alone 16, or 100 times more samples) can be totally infeasible. We henceforth use the term optimal to refer to protocols that are exactly optimal, and will specify when a protocol is merely asymptotically optimal.

\section{Prior work}
\label{priorwork}

We here detail a selection of works critical to the development of LDP distribution estimation, each optimal in some parameter regime. Among these optimal protocols, we further distinguish between those that also minimise communication cost. We will say that a protocol achieves optimal communication if it uses $O(\log\ab)$ bits, which in this work implies that $\oab=O(\ab)$. 
As we will see, many historically optimal protocols have implicitly used combinatorial block designs, a connection made explicit by Park et al.~\cite{park2024} who recently demonstrated that protocols based on BIBDs achieve minimax optimality. Specifically, they introduce the following family of local randomisers.

\begin{definition}[Block Design Randomised Response~{\cite[Def. 5]{park2024}}]
	\label{blockdesignrr}
	For a $\bvr$-configuration $\mathcal{R}$ with point set $\dmain = [\ab]$ and block set $\{\cB_1,\ldots,\cB_\oab\}$, each of size $\ssz$, let the output range $\range = [\oab]$. Define $S(j)\subset \range$ to be the set of blocks containing point $j \in \dmain$, with $|S(j)| = \reps$. Then the local randomiser $\Pmech$ defined by
	\begin{equation*}
		\Pmech_{ij} = \begin{cases}
			\frac{e^\priv}{\reps(e^\priv - 1) + \oab}, & \text{if }i\in S(j), \text{ and} \\
			\frac{1}{\reps(e^\priv - 1) + \oab} &\text{otherwise,}
		\end{cases}
	\end{equation*}
	achieves optimal error when $\ssz$ is the optimal subset size as defined in~\cref{chaisubsetsize}. If $\mathcal{R}$ is a \textit{symmetrical} BIBD, then it also achieves optimal communication.
\end{definition}

Let $\Pmech$ be a local randomiser of the form defined above, let $A$ be the incidence matrix of a BIBD. \Cref{blockdesignrr} implies that the TPM of $\Pmech$ may be written in the form $p(A(e^\priv - 1) + J')$, where $p=1/(r(e^\priv - 1) + \oab)$ and $J'$ is the all-ones matrix of size $\oab\times\ab$. When we say that a local randomiser $\Pmech$ uses a BIBD, we mean that its TPM is of this form, with $A$ being the incidence matrix of a BIBD. To communicate the importance of these constructions for locally private distribution estimation, we now detail a selection of historically important protocols that are each optimal in some parameter regime.

Binary randomised response~\cite{warner1965} (2-RR) predates differential privacy\footnote{Generalisation of randomised response were considered by statisticians for some time before differential privacy was conceived. We direct interested readers to the short book~\cite{chaudhuri1988}.} and existed originally as a simple procedure. Given a yes-no question, flip a coin; if the coin is heads then answer truthfully, if it is tails then flip the coin again and answer ``yes'' if heads or ``no'' if tails. Under differential privacy, we consider it differently: flip a weighted coin that has heads probability $e^\priv/(e^\priv + 1)$, if it is heads, then answer truthfully, otherwise answer the opposite. 

The natural extension of binary randomised response, called $\ab$-randomised response ($\ab$-RR) was likely the first extension of binary randomised response to domains larger than 2. This protocol simply returns the true response with probability $e^\priv/(e^\priv + \ab - 1)$. It is apparent that as $\ab$ grows large relative to $\priv$, the induced output distribution becomes close to uniform, regardless of the input. For this reason, $\ab$-RR is only optimal when $\priv>\log\ab$. This protocol correspond to the trivial design with $A=I_\ab$.

An early work detailing necessary conditions for optimality demonstrated that the TPM of an optimal LDP distribution estimation protocol must have a binary structure, only containing ``small'' and ``large'' elements such that the ratio of large over small equals $e^\priv$~\cite{kairouz2016}. Their methodology, optimising for the mutual information between input and output distributions of a protocol inspired the development of ``Subset Selection'', or $\ssz$-subset protocols~\cite{wang2016,ye2018a}. 

On input $x$ Subset Selection protocols return, with a probability greater than $1/2$ depending on the privacy parameter $\priv$, a $\ssz$-subset of $\dmain$ containing $x$. Otherwise it returns a uniformly random $\ssz$-subset of $\dmain\setminus\{x\}$~\cite{wang2016,ye2018a}. This protocol corresponds to the trivial design with incidence matrix $A$ with $\oab=\binom{\ab}{\ssz}$ rows, each representing a distinct $\ssz$-subset of $\dmain$. An optimal choice of $\ssz$ was found by maximising mutual information between a uniform input distribution and the induced output distribution~\cite{wang2016}. In the extreme, when $\priv>\log\ab$, we recover the $\ab$-RR protocol. The communication cost of these protocols is the minimum of $\ssz\lceil\log_2\ab\rceil$ and $\ab$ bits, as we can either specify a row by the indices of its large elements, or by a bit-string encoding their positions.
An important reference for our purposes will be~\cite{chai2023}, who analysed the Subset Selection protocol through spectral properties of its TPM.

Hadamard Response (HR)~\cite{acharya2019a} achieved optimal communication with $\oab=\ab$ in the high privacy (low $\priv$) regime by using Hadamard matrices (of Sylvester's construction with the first row and column removed) to define the position of large elements\footnote{We remark that Hadamard Response actually has a much older history in Randomised Response procedures, first dating to 1979~\cite{raghavarao1979} as found in~\cite[Chapter 5]{padgett2005}.}. These matrices correspond to the symmetric BIBDs known as Hadamard configurations. This makes Hadamard Response the first non-trivial application of SBIBDs to LDP distribution estimation. Hadamard matrices, by their nature, only exist with subsets of size $\ssz\approx\ab/2$~\cite{sylvester1908}, and as such cannot always satisfy the optimal subset size derived in earlier works. To resolve this, the authors introduced a block-diagonal generalisation of their protocol which interpolated between HR and $\ab$-RR. This generalisation performs poorly in practice~\cite{feldman2022a,canonne2025}. The need for a protocol combining HR's optimal communication cost, and Subset Selection's optimal error motivated the introduction of Projective Geometry Response.

Projective Geometry Response (PGR)~\cite{feldman2022a} uses the incidence structure of projective planes over a finite field $\mathbb{F}_\pp^t$. Each input ``point'' is mapped to a distinct hyperplane which forms a ``line'' of high-probability elements. This incidence structure corresponds to that of the point-hyperplane SBIBD, or the SBIBD given by Singer difference sets~\cite{singer1938,colbourn2006,hall1986}. Selecting $\pp\approx e^\priv+1$ minimises the variance of the protocol, and results in a subset size (the number of points incident to each line) of $\ssz\approx\ab/(e^\priv + 1)$. This protocol generalises well in the case when $1<\priv<\log\ab$. However, these set structures only exist when $\ab$ is of the form $\frac{\pp^t - 1}{\pp - 1}$, and as such, can require rounding of the input size from $\ab$ to $O(e^\priv\ab)$.

As we have seen, the general definition of block design randomised response by Park et al.~\cite{park2024} subsumes randomised response, subset selection, HR, and PGR. The authors further demonstrate that all protocols based on block designs are exactly minimax optimal when $\ssz\approx\ab/(e^\priv + 1)$. It has remained open whether there exist protocols achieving the same performance based on other, perhaps more practical, constructions. Our main contribution is to establish that \emph{block designs are in fact necessary for optimality}.

\section{Optimal local protocols}
\label{secoptimallocalpros}
We now detail in more depth the necessary conditions for optimality of LDP distribution estimation protocols. \Cref{optimalldpprops} details a list of properties that all optimal LDP protocols are known to satisfy. We state in full the properties whose details are neccessary in establishing our main result.

\begin{theorem}
	\label{optimalldpprops}
		The following is true of optimal LDP distribution estimation algorithms:
	\begin{enumerate}
		\item\label{optbinary} Their transition probability matrix contains only two distinct values, call them ``small'' and ``large'', 
		\item\label{optratio} the ratio between these elements is in $\{e^{-\priv},1,e^\priv\}$,
		\item\label{optosize} and the output size $\oab=|\range|$ is at most $\ab=|\dmain|$~\cite[Theorem 2]{kairouz2016}.
		\item\label{optssize} Each row of $\Pmech$ contains exactly $\ssz$ large elements, where $\ssz$ is either the floor or ceiling of $\ab/(e^\priv + 1)$~\cite{wang2016,ye2018a,chai2023}.
		\item\label{opteigen} The Gram matrix $\Pmech'\Pmech$ has exactly two eigenvalues with multiplicity $1$ and $\ab-1$.~\cite{chai2023}.
	\end{enumerate}
	We will make the further assumption that the TPM of optimal local randomisers have no repeated rows, as any two equal rows can trivially be merged into one.
\end{theorem}

For a local randomiser $\Pmech$ and the induced distribution $\freqs$ given by~\cref{induceddist}, let the optimal unbiased estimator for $\dist$ be $L$ such that $\bEE{L\freqs}=\dist$. This implies that $\debias\Pmech=I$. 
\begin{lemma}[{\cite[Prop.~2.1]{chai2023}}]
\label{minimaxrisk}
Let $D_\freqs$ be the matrix of order $\oab$ with $\freqs$ on the diagonal. The minimax risk of learning under a locally private protocol $\Pmech$ is lower bounded by
	\begin{equation}
		\ns\bEE{\norm{\hat\dist - \dist}_2^2}\geq \tr(\Pmech'D_\freqs^{-1}\Pmech)^{-1} - \norm{\dist}_2^2,
	\end{equation}
	and the lower bound is reached when 
	\begin{equation}
		L = (\Pmech' D_\freqs^{-1}\Pmech)^{-1}\Pmech'D_\freqs^{-1}.
	\end{equation}
\end{lemma}

\Cref{minimaxrisk} implies that optimality depends only on the transition probability matrix of the protocol, not on the debiasing function, and that the optimal protocol is the one that maximises the trace $\tr(\Pmech'D_\freqs^{-1}\Pmech)$. 

\begin{lemma}[{Optimal subset size~\cite[Lemma~3.1]{chai2023}, \cite{wang2016}}]
\label{chaisubsetsize}
	For $\priv>0$ and $\ab>2$, let 
	\begin{equation}
		f(x) = \frac{\ab^2(x (e^\priv)^2 - x)}{(x(e^\priv - 1) + \ab)^2},\quad x \geq 0
	\end{equation}
	and 
	\begin{equation}
		q = \begin{cases}
			\left\lfloor  \frac{\ab}{1+ e^\priv } \right\rfloor,  &\text{if } f\left( \left\lfloor  \frac{\ab}{1+e^\priv}  \right\rfloor  \right) \geq f\left( \left\lceil  \frac{\ab}{1+ e^\priv }  \right\rceil  \right)\\
			&\qquad\text{ and } \left\lfloor  \frac{\ab}{1+ e^\priv }  \right\rfloor \geq 1 \\ \left\lceil  \frac{\ab}{1+ e^\priv }  \right\rceil, &\text{otherwise.}
		\end{cases}
	\end{equation}
	Then for all protocols satisfying~\cref{optimalldpprops}~\cref{optbinary,optratio} having no two rows proportional to each other, then
	\begin{equation}
		\tr(\Pmech'D_\freqs^{-1}Q) \leq f(q).
	\end{equation}
	With equality if each row of $\Pmech$ contains exactly $q$ large values.
\end{lemma}

\Cref{chaisubsetsize} demonstrates, surprisingly, that the optimal trace is independent of the input distribution, and combined with~\cref{minimaxrisk} we get that the uniform distribution, which minimises $\norm{\mu}_2^2$, is the worst case.

\begin{lemma}[{\cite[Lemmas 3.2 and 3.3]{chai2023}}]
\label{chaimintrace}
	A lower bound on the trace $\tr(\Pmech'D_\freqs^{-1}\Pmech)^{-1}$ is
	\begin{equation}
		\frac{(\ab-1)^2}{f(q) - \ab} + \frac{1}{\ab},
	\end{equation}
	and it is attained if and only if the eigenvalues of $\Pmech'D^{-1}_\freqs\Pmech$~are
	\begin{align*}
		\eigen_{i} &= \frac{f(q)-\ab}{\ab-1},\tag{$i=1,2,\dots,\ab-1$}\intertext{and}
		\eigen_{\ab} &= \ab, 
	\end{align*}
	In particular this implies that 
	\begin{equation}
		\Pmech'D_\freqs^{-1}\Pmech = a_qI + b_qJ
	\end{equation}
	for $a_q = \frac{f(q)-\ab}{\ab-1}$ and $b_q = 1-a_q/\ab$.
\end{lemma}

These criteria are appealing in that they provide \textit{exact} criteria for achieving a lower bound that is itself exact. If a protocol achieves these bounds, there remains no ambiguity about whether it will perform poorly due to constant factors.

Finally, we demonstrate that~\cref{optbinary,optssize}~of~\cref{optimalldpprops} implies that on a uniform input distribution, the induced distribution $\freqs$ is also uniform.

\begin{lemma}
\label{uniformdist}
	Let $U_\ab$ be the discrete uniform distribution on $\ab$ elements, and $\Pmech$ be a local randomiser satisfying~\cref{optbinary,optssize}~of~\cref{optimalldpprops}. Then the induced distribution,
	\begin{equation*}
		\freqs=\Pmech U_\ab=U_\oab,
	\end{equation*}
	is the discrete uniform distribution on $\oab$ elements.
\end{lemma}
\begin{IEEEproof}
	The constant row sum of $\Pmech$ ensures that the $i$th component of $\freqs$ is given by
	\begin{align*}
		\freqs_i &= \frac{1}{\ab}\sum_{j=1}^\ab \Pmech_{ij}=\frac{1}{\ab}(\ssz\Lel + (\ab-\ssz)\Sel).
	\end{align*}
	As this quantity is a constant independent of $i$ and must sum to one, the induced distribution is uniform.
\end{IEEEproof}

\section{Necessity of designs}
\label{optimaldesigns}

We now proceed with the main theorem of this paper, demonstrating that it is \emph{necessary} that an optimal local randomiser be derived from a BIBD. While Park et al.~\cite{park2024} showed that BIBD-based protocols achieve optimality, we prove the converse: if a protocol achieves optimality for a given $\priv$ and $\ab$, then its transition probability matrix must be derived from a BIBD with $\ssz\approx\ab/(e^\priv + 1)$.

\begin{theorem}
\label{optdpimpliesbibd}
	Let $\Pmech$ be a locally private protocol that achieves~\cref{optbinary,optratio,optssize,opteigen}~of~\cref{optimalldpprops}. Then the transition probability matrix of $\Pmech$ must be of the form
	\begin{equation}
		\Pmech = p(A(e^\priv - 1) + J'),
	\end{equation}
	where $A$ is a $\bvr$-configuration and $J'$ is the $\oab\times\ab$ all-ones matrix. Furthermore, if $\Pmech$ satisfies \cref{optosize}, then $A$ must be a SBIBD.
\end{theorem}
\begin{IEEEproof}
Let $\Pmech$ be the transition probability matrix of a locally private protocol that satisfies~\cref{optbinary,optratio,optssize,opteigen}~of~\cref{optimalldpprops}. We disregard~\cref{optosize} in the interest of proving a more general result. \Cref{optbinary} implies that $\Pmech$ can be written as $Q=A(\Lel-\Sel) + \Sel J'$, for some zero-one matrix $A$ that defines the position of the large entries. \Cref{optssize} explicitly gives $A$ constant row sum. Recall that $J$ is the all-ones matrix of order $\ab$, $J'$ is the rectangular all-ones matrix of size $\oab\times\ab$ and $J''$ is its transpose.

By~\cref{minimaxrisk,chaisubsetsize} the protocol that achieves the best error on the uniform distribution is minimax optimal. By~\cref{uniformdist} we have that the induced distribution $\freqs$ is also uniform. This implies that the matrix $D_\freqs$ is the square matrix of order $\oab$ with $1/\oab$ on the diagonal. Therefore $D_\freqs^{-1}=\oab I_\oab$ is simply a scalar multiplication by $\oab$ and can be factored out.

We know by~\cref{chaimintrace} that
\begin{equation*}
	\Pmech'D_\freqs^{-1}\Pmech = a_qI + b_qJ.
\end{equation*}
for $a_q = \frac{f(q)-\ab}{\ab-1}$ and $b_q = 1-a_q/\ab$, where $q$ is the optimal subset size defined in~\cref{chaisubsetsize}. 

We will see that to equate both expressions $A$ must have some constant column sum $\reps$. Let $\vec{r}=(r_1,\dots,r_\ab)$ be the column sums of $A$, and let $D_{\vec{r}}$ be the matrix with $\vec{r}$ on its diagonal. Clearly $A'J=D_{\vec{r}}J$ and $J''A=JD_{\vec{r}}$. To be in the form above, we must have $A'J'+J''A=tJ$ for some constant $t$. This can only be the case when $\vec{r}=(r,r,\dots,r)$ is a constant vector, in which case $t=2r$.

Furthermore, the equation given by~\cref{chaimintrace} has only terms in $I$ and $J$, we must have $A'A=c_1I+c_2J$ for some constants $c_1,c_2$. We will solve the equality for $c_1$ and $c_2$, demonstrating that they equal exactly $\reps-\sint$ and $\sint$ respectively, so that $A$ is the incidence matrix of a BIBD.

Using the column sum constraint, we can simplify further. Let $p= 1/{(\reps(e^\priv - 1) + \oab)}$ be the normalisation factor, replacing $\Lel$ with $pe^\priv$, and $\Sel$ with $p$ in $\Pmech$. 

Counting the number of large elements in $A$ in two ways, we can see that each row contains $\ssz$ large elements, and there are $\oab$ rows, giving that there are $\ssz\oab$ large elements. Then, as each column contains $\reps$ large elements, and there are $\ab$ columns, we have that the total number of large elements is also $\reps\ab$. Equating these two expressions gives us that $\ssz\oab=\reps\ab$, allowing us to make several useful substitutions.

For one, we can show that $p=\frac{1}{b+r(e^\priv - 1)}=\frac{\ab}{\oab(\ab+\ssz(e^\priv - 1))}$.
Letting $p_0=\frac{1}{\ab+\ssz(e^\priv - 1)}$, we can see that $p=\frac{\ab}{\oab}p_0$.

We proceed to solve for $c_1$ and $c_2$ separately, by looking at the coefficients of $I$ and $J$ respectively. First, it is helpful simplify $a_q$ and $b_q$. We can see that
\begin{align*}
	a_{q} & =\frac{f(q)-\ab}{\ab-1} \\
		& = \frac{\ab \ssz(\ab-\ssz){ (e^{ \priv }-1)^2 }}{(\ab-1){ (\ab+\ssz(e^{ \priv }-1))^2 }}  \\
		& =p_{0}^2(e^{ \priv }-1)^2 \frac{\ab\ssz(\ab-\ssz)}{\ab-1},\\
	b_{q}  & =1- \frac{a_{q}}{\ab}  \\
		& = \frac{(\ab-1)(\ab+\ssz(e^{ \priv }-1))^2-\ssz(\ab-\ssz)(e^{ \priv }-1)^2}{(\ab-1)(\ab+\ssz(e^{ \priv }-1))^2}  \\
		& = p_{0}^2 \left( (\ab+\ssz(e^{ \priv }-1))^2- \frac{\ssz(\ab-\ssz)(e^{ \priv }-1)^2}{(\ab-1)}  \right) \\
		& =p_{0}^2\left( \left( \ssz^2- \frac{\ssz(\ab-\ssz)}{\ab-1} \right)(e^{ \priv }-1)^2  +2\ab\ssz(e^{ \priv }-1) + \ab^2 \right)
\end{align*}

Then, looking again at $\Pmech'D_{\freqs}^{-1}\Pmech$ and substituting $\ssz=\reps\ab/\oab$, we get
\begin{align*}
\Pmech'D^{-1}_{\freqs}\Pmech  & = \oab p^2(A(e^{ \priv }-1)+J')(A'(e^{ \priv }-1)+J'') \\
 & = \oab p^2((c_{1}(e^{ \priv }-1)^2I + (c_{2}(e^{ \priv }-1)^2+2r(e^{ \priv }-1)+\oab)J) \\
 & = p_{0}^2\bigg( \frac{\ab^2}{\oab}c_{1}(e^{ \priv }-1)^2I + \left( \frac{\ab^2}{\oab}c_{2}(e^{ \priv  }-1 )^2 + 2\ab\ssz(e^{ \priv }-1) + \ab^2 \right)J \bigg) 
\end{align*}
To solve for $c_{1}$, we equate the coefficients of $I$ on each side and substitute $\reps={\oab\ssz}/{\ab}$
\begin{align*}
	\frac{\ab^2}{\oab}p_{0}^2c_{1}(e^{ \priv }-1)^2   & = p_{0}^2(e^{ \priv }-1)^2 \frac{\ab\ssz(\ab-\ssz)}{\ab-1} \\
c_{1} & = \frac{\oab\ssz(\ab-\ssz)}{\ab(\ab-1)} \\
 & = \frac{\reps(\ab-\ssz)}{\ab-1} \\
 & = \reps - \frac{\reps(\ssz-1)}{\ab-1}
\end{align*}
Solving for $c_{2}$, we repeat the same process for the coefficients of $J$. Cancelling $p_{0}^2$ and the equal additive terms on each side, we are left with
\begin{align*}
	p_0^2\left( \frac{\ab^2}{\oab}c_{2}(e^{ \priv  }-1 )^2 + 2\ab\ssz(e^{ \priv }-1) + \ab^2 \right) & = p_{0}^2\left( \left( \ssz^2- \frac{\ssz(\ab-\ssz)}{\ab-1} \right)(e^{ \priv }-1)^2  +2\ab\ssz(e^{ \priv }-1) + \ab^2 \right) \\
c_{2} & = \frac{\oab\ssz^2}{\ab^2}- \frac{\oab\ssz(\ab-\ssz)}{\ab^2(\ab-1)} \\
 &=\frac{\reps\ssz}{\ab}- \frac{\reps(\ab-\ssz)}{\ab(\ab-1)} \\
 & =\frac{\reps(\ssz-1)}{\ab-1}.
\end{align*}
This finally gives us
\begin{equation*}
	A'A=\left( \reps- \frac{\reps(\ssz-1)}{\ab-1} \right)I + \frac{\reps(\ssz-1)}{\ab-1}J.
\end{equation*}
Now, given that $A$ is a zero-one $\oab\times\ab$ matrix, and we have that $A'A=(\reps-c_{2})I+c_{2}J$ for $c_{2}$ above, the diagonal entries of $A'A$ are exactly $\reps$, the number of ones in each column. For the off-diagonal entries, $i\neq j$, we have $(A'A)_{ij}=\sum_{t=1}^\ab A_{ti}A_{tj}$, which (since $A$ is zero-one) counts exactly the number of rows where columns $i$ and $j$ have a one. This value being constant necessitates that any two columns share exactly $\sint= \frac{\reps(\ssz-1)}{\ab-1}$ positions in common, demonstrating that 
\begin{align*}
AJ & =\ssz J', \\
A'A & =(\reps-\sint)I+\sint J.
\end{align*}
By~\cref{defbibd} we have that $A$ is a $\bvr$-configuration. Furthermore, Fischer's inequality states that a BIBD only exists with $\oab\geq\ab$~\cite[p. 99]{ryser1963}, combined with~\cref{optosize} we have that $\ab\leq\oab$, and $A$ must be a $\vkl$-configuration with $\oab=\ab$.
\end{IEEEproof}

\section{Future work}
\label{futurework}

As shown by~\cite{park2024}, the Block Design Randomised Response protocols are optimal. Our work demonstrates that these are in fact the \emph{only} optimal protocols. While Park et al. provide a thorough survey of BIBDs for locally private distribution estimation, we point interested readers to the 21 families of SBIBDs cited in~\cite[Section II.6.8]{colbourn2006}, the BIBDs with $\sint=1$ that can be instantiated for all $\ab$ and all values of $\ssz\leq 5$~\cite{stinson2004,colbourn2001}, the list of implementations in~\cite[Reference Manual, Combinatorics, Balanced Incomplete Block Designs]{sagemath}, and the field of coding theory~\cite{assmus1992, xu2022}.

We demonstrate optimality with respect to $\ell_2^2$ error, and prior work suggests that these protocols also maximise mutual information; however, there may be other norms of interest, for which different constructions could be optimal.%

While we have shown necessary conditions for optimal LDP distribution estimation, there are many other statistical queries which are not so well understood. We hope that future work will reveal the structure of optimal protocols for these queries.
\bibliographystyle{IEEEtran}
\bibliography{IEEEabrv, references.bib}

\end{document}